\documentclass[pra,twocolumn,superscriptaddress,nofootinbib,noshowpacs,preprintnumbers,longbibliography,floatfix]{revtex4-2}
\usepackage[utf8]{inputenc}
\usepackage[pdftex]{graphicx}
\usepackage{float}
\usepackage{amssymb}
\usepackage{amsmath}  
\usepackage{dsfont}
\usepackage{array}
\usepackage{bm,fixmath}
\usepackage{mathrsfs}
\usepackage{pifont}
\usepackage{multirow}
\usepackage{upgreek}
\usepackage{xcolor}
\usepackage{bm}
\usepackage{bbm}
\usepackage{comment}
\usepackage[pdftex,
            pdftitle={Simulating the Flight Gate Assignment Problem on a Trapped Ion Quantum Computer},
            pdfauthor={Authors}, 
            bookmarks,
            colorlinks,
            linkcolor=myblue,
            citecolor=mymagenta,
            menucolor=black,
            urlcolor=myblue,
            plainpages=false,
            pdfpagelabels,
            hypertexnames=false]{hyperref}
\usepackage[braket,qm]{qcircuit}
\usepackage{cleveref}
\usepackage{bm}
\definecolor{mymagenta}{RGB}{200, 0, 100}
\definecolor{myblue}{RGB}{45, 48, 146}

\begin{document}
\title{Simulating the flight gate assignment problem on a trapped ion quantum computer}

\author{Yahui Chai}
\affiliation{
CQTA, Deutsches Elektronen-Synchrotron DESY, Platanenallee 6, 15738 Zeuthen, Germany
}

\author{Evgeny Epifanovsky}
\affiliation{
IonQ Inc, 4505 Campus Dr, College Park, Maryland 20740, USA
}

\author{Karl Jansen}
\affiliation{
CQTA, Deutsches Elektronen-Synchrotron DESY, Platanenallee 6, 15738 Zeuthen, Germany
}
\affiliation{
Computation-Based Science and Technology Research Center, The Cyprus Institute, 20 Kavafi Street,
2121 Nicosia, Cyprus
}

\author{Ananth Kaushik}
\affiliation{
IonQ Inc, 4505 Campus Dr, College Park, Maryland 20740, USA
}

\author{Stefan Kühn}
\affiliation{
CQTA, Deutsches Elektronen-Synchrotron DESY, Platanenallee 6, 15738 Zeuthen, Germany
}

\date{\today}

\begin{abstract}
    We study the flight gate assignment problem on IonQ's Aria trapped ion quantum computer using the variational quantum eigensolver. Utilizing the conditional value at risk as an aggregation function, we demonstrate that current trapped ion quantum hardware is able to obtain good solutions for this combinatorial optimization problem with high probability. In particular, we run the full variational quantum eigensolver for small instances and we perform inference runs for larger systems, demonstrating that current and near-future quantum hardware is suitable for addressing combinatorial optimization problems.
\end{abstract}

\maketitle

\section{Introduction\label{sec:intro}}
Combinatorial optimization problems are of paramount importance for real-world applications ranging from finance~\cite{Merton1973,Juarna2017} to logistics~\cite{Sbihi2010} to telecommunication~\cite{Resende2003}. A particularly important subclass of combinatorial optimization problems are quadratic assignment problems. Originally introduced in the context of economics~\cite{Koopmans1957}, these problems deal with assigning a set of facilities to a set of locations in such a way that the cost is minimized. A particular instance of a quadratic assignment problem is the flight gate assignment (FGA) problem~\cite{Dorndorf2007}, whose objective is to find an assignment of flights to the gates of an airport, such that the total transit time of all passengers in the airport is minimized. Quadratic assignment problems are, in general, known to be NP-hard~\cite{Finke1987, Garey1979}. Thus, in practice, it is often impossible to compute the full solution, and one has to resort to heuristic algorithms that have been empirically proven to provide a good approximation to the solution for the problem at hand.  

Quantum computing might offer an alternative approach to such tasks. While NP-hard problems can, in general, also not be solved efficiently on a quantum computers~\cite{Nielsen2010}, first studies suggest that it might be possible to find good approximations for the solution of combinatorial optimization problems using quantum devices~\cite{Stollenwerk2018,Mohammadbagherpoor2021,Liu2022,Chai2023}. In particular, variational quantum algorithms~\cite{Moll2018,Cerezo2021} seem to be promising candidates for addressing combinatorial optimization problems on current and near-future noisy intermediate-scale quantum (NISQ) devices. These algorithms utilize a parametric quantum circuit as a variational ansatz whose parameters can be trained to extremize a given cost function. Prominent examples are the quantum adiabatic optimization algorithm~\cite{Farhi2014} and the variational quantum eigensolver (VQE)~\cite{Peruzzo2014}. In both approaches the problem is encoded into a (classical) Hamiltonian whose ground state represents the solution of the problem. Thus, optimizing the parameters in the ansatz such that the energy expectation value is minimal, the quantum circuits encodes a candidate for the solution of the problem. While the quantum adiabatic optimization algorithm uses an ansatz that is inspired by adiabatically transforming an easily preparable initial state into the solution, it typically requires circuit depths that are challenging for current quantum devices. In contrast, VQE does not pose any restrictions on the choice of ansatz circuit, which in conjunction with recent improvements~\cite{Barkoutsos2020,Amaro2022} renders it a promising candidate for addressing combinatorial optimization problems with NISQ devices.

In this paper, we explore the performance of VQE for the flight gate assignment (FGA) problem~\cite{Dorndorf2007} on a quantum device using the approach from Ref.~\cite{Chai2023}. While the FGA problem has been addressed previously on superconducting hardware utilizing a similar approach~\cite{Mohammadbagherpoor2021}, we focus on trapped ion quantum hardware and use IonQ's Aria device. Studying the problem on a variety of quantum architectures helps understand the strengths and weaknesses of each modality. For example, trapped ion quantum computers permit all to all qubit connectivity while superconducting architectures are limited to a fixed topology. This affects the types of quantum circuits that could be implemented without the overhead of swap gates. In the following, we demonstrate the full VQE for smaller problem instances and we perform inference runs for large problem instances with up to 18 qubits. 

The paper is structured as follows. In Sec.~\ref{sec:model}, we briefly summarize the problem and we discuss the binary encoding from Ref.~\cite{Chai2023} that we are using in our simulations. Subsequently, we introduce the setup for our hardware runs in Sec.~\ref{sec:setup}. Section~\ref{sec:results} contains the results obtained on  IonQ's Aria trapped ion quantum computer. Finally, we conclude in Sec.~\ref{sec:conclusion}.

\section{The flight gate assignment problem and its encoding into qubits\label{sec:model}}

The objective of the FGA problem is to find an optimal assignment of flights to the gates at an airport. Although one can think about multiple metrics for judging the quality of an assignment, we focus on minimizing the total transit time of all passengers in the airport~\cite{Kim2017}. Here we consider three kinds of passengers: arriving passengers coming with an inbound flight and leaving the airport from their arrival gate, departing passengers that enter the airport through the security check and having to walk to their departure gate, and transit passengers arriving at one gate and having to walk to the gate of their connecting flight. In order to model this problem mathematically, and to encode it into qubits, we start from a formulation as a quadratic unconstrained binary optimization (QUBO) problem with linear constraints. The cost function of the QUBO problem can then be translated into a Hamiltonian that can be addressed on the quantum device. In the rest of the paragraph we briefly summarize these steps following Ref.~\cite{Chai2023}.

Considering a set of flights $F$ and a set of gates $G$, the FGA problem can be represented as a QUBO by using ${|F| \times |G|}$ binary variables $x_{i\alpha} \in \{0, 1\}$, which take the value of 1 if and only if flight $i\in F$ is assigned to gate $\alpha\in G$. The total time the passengers spend in the airport given an assignment $x\in\mathds{Z}_2^{|F|\times|G|}$ then reads
\begin{align}
    T(x) = T^a(x) + T^d(x) + T^t(x),
    \label{eq:total_time}
\end{align}
where $T^a(x)$, $T^d(x)$ , and $T^t(x)$ represent the total times for the arriving, departing and transit passengers. They are given by
\begin{align}
    T^{a/d}(x) &= \sum_{i,\alpha} n_i^{a/d}t_\alpha^{a/d}x_{i\alpha},\label{eq:t_ad}\\
    T^{t}(x) &= \sum_{i,j,\alpha,\beta} n_{ij}t_{\alpha\beta}x_{i\alpha}x_{j\beta},\label{eq:t_t}    
\end{align}
where $t_\alpha^{a/d}$ is the time it takes for the arriving/departing passengers to walk from/to gate $\alpha$, $t_{\alpha\beta}$ is the time it takes the transit passengers to walk from gate $\alpha$ to gate $\beta$, $n_i^{a/d}$ are the number of passengers arriving/departing with flight $i$, and $n_{ij}$ is the number of passengers transferring from flight $i$ to $j$.

In addition, one has to take into account two linear constraints. On the one hand, each flight can only be assigned to a single gate at a time. This can be ensured by imposing
\begin{align}
    \forall i \in F \quad \sum_\alpha x_{i\alpha} = 1.
    \label{eq:constraint1}
\end{align}
On the other hand, two flights cannot be scheduled at the same gate within a certain time span, which comprises the time a plane spends at the gate and some buffer time $t^\text{buffer}$ between the departure of the first flight and the arrival of the second flight. This can be expressed mathematically as
\begin{align}
    \forall \alpha\in G \ \text{and}\  \forall (i,j)\in O\quad  x_{i\alpha}\times x_{j\alpha} = 0,
    \label{eq:constraint2}
\end{align}
where $O$ is the set of overlapping flight pairs,
\begin{align}
    O = \{ (i,j) \in F\times F : t_i^\text{in} < t_j^\text{in} < t_i^\text{out} + t^\text{buffer} \}.
\end{align}
with $t_i^\text{in}$ the arrival time of flight $i$ and $t_j^\text{out}$ the departure time of flight $j$. For the rest of the paper, we refer to assignments fulfilling both of these constrains as feasible ones.

In principle, the QUBO formulation described above can be directly turned into a Hamiltonian by substituting the decision variables $x_{i\alpha}$ in the cost function with the operators $(\mathds{1} - Z_{i\alpha})/2$, where $Z_{i\alpha}$ is the Pauli $Z$-matrix acting on the qubit corresponding to the decision variable $x_{i\alpha}$. The resulting Hamiltonian would be diagonal in the computational basis, with the different solutions being encoded in the the computational basis states $\ket{x}$, $x=0,\dots,2^{|F|\times |G|}-1$. This approach requires $N = |F| \times |G|$ qubits, however due to the constraints in Eqs.~\eqref{eq:constraint1} and \eqref{eq:constraint2} only an exponentially small fraction of states in the Hilbert space would correspond to feasible assignments~\cite{Chai2023}. 

A more resource efficient encoding into qubits can be obtained by constructing a Hamiltonian that directly confined to the subspace that fulfills Eq.~\eqref{eq:constraint1}. The constraint enforces that for each $i\in F$ there can be at most a single decision variable $x_{i\alpha}$, $\alpha = 0,\dots, |G|-1$ nonzero. The feasible assignments for each flight can thus be represented by $M = \lceil \log_2(|G|) \rceil $ (qu)bits. In general, the number of gates is not going to be a power of 2, and we choose to map the elements in $G$ cyclically to the basis states $\ket{\alpha'}$, $z=0,\dots,2^M-1$, where the state $\ket{\alpha'}$ corresponds to the gate $\alpha = \alpha' \mod |G| \in G$. Using this approach, we need a total of $Q = M \times |F|$ qubits to encode the problem on a digital quantum computer, which is a lot less than for the QUBO formulation discussed above. Moreover, the cyclic  mapping generally leads to multiple degenerate solution states, which enhances the possibility of finding an optimal solution, similar to degeneracy engineering~\cite{Anschuetz2022}.

Using this encoding, the individual contributions to the cost function in Eqs.~\eqref{eq:t_ad} and \eqref{eq:t_t} can be expressed as Hamiltonians~\cite{Chai2023}
\begin{align*}
    H^{a/d} &= \sum_{i} \sum_{\alpha^{\prime} = 0}^{2^M-1} n_i^{a/d} t_{\alpha}^{a/d} P_i(\alpha^{\prime}), \\
    H^{t} &= \sum_{ij} \sum_{\alpha^{\prime} \beta^{\prime} = 0}^{2^M-1} n_{ij} t_{\alpha\beta}  P_i(\alpha^{\prime}) P_j(\beta^{\prime}).
\end{align*}
In the expression above  $\alpha$,$ \beta$ refer to the actual gate indices obtained from the cyclically mapped ones $\alpha'$ and $\beta'$, and $P_j(\alpha')$ are projection operators given by
\begin{equation}
    \begin{aligned}
        P_i(\alpha^{\prime}) &= \op{\alpha^{\prime}}{\alpha^{\prime}}_i = \op{z_0 \cdots z_{M-1}}{z_0 \cdots z_{M-1}}_i \\
        &= \left( \op{z_0}{z_0}  \otimes \dots \otimes \op{z_{M-1}}{z_{M-1}} \right)_i.
    \end{aligned}
\end{equation}
Here $z_0 \cdots z_{M-1}$ denotes the bit string for the binary representation of $\alpha'$, and the index $i$ indicates the set of qubits related to flight $i$. The individual terms can be easily represented in terms of Pauli matrices. Choosing a linear ordering of the qubits, they can be expressed as 
\begin{align}
    \op{z_k}{z_k}_i = \frac{1}{2}\left(\mathds{1} + (-1)^{z_k}Z_{i\times M + k}\right).
\end{align}

The formulation above automatically incorporates the constraint that only a single flight can be assigned to a gate at a time. In addition, the constraint from Eq.~\eqref{eq:constraint2} has to be enforced, too. Here we chose to do so by adding a positive semidefinite penalty term $H^O$ to the overall Hamiltonian
\begin{align}
    H^O &= \sum_{(i,j)\in O} \sum_{\alpha^{\prime} \beta^{\prime}= 0}^{2^M-1} \delta_{\alpha \beta} P_i(\alpha^{\prime}) P_j(\beta^{\prime}).
\end{align}
Considering the problem Hamiltonian and the penalty term, the final Hamiltonian reads
\begin{align}
    H = H^a + H^d + H^t + \lambda H^O,
    \label{eq:Hamiltonian}
\end{align}
where $\lambda>0$ is constant that has to be chosen large enough to ensure that the solution is in the subspace fulfilling the constraint.

\section{Experimental setup\label{sec:setup}}
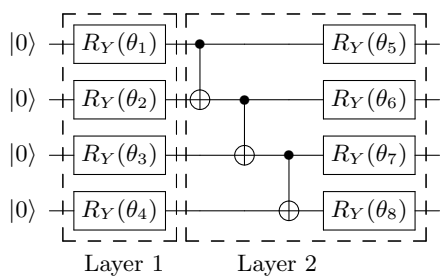
\begin{figure}[htp!]
    \centering
    \begin{align*}
        \Qcircuit @C=1em @R=.7em {
        \lstick{\ket{0}} & \gate{R_Y(\theta_1)} & \ctrl{1}                             & \qw      & \qw      & \gate{R_Y(\theta_5)} & \qw\\
        \lstick{\ket{0}} & \gate{R_Y(\theta_2)} & \targ                                & \ctrl{1} & \qw      & \gate{R_Y(\theta_6)} & \qw \\
        \lstick{\ket{0}} & \gate{R_Y(\theta_3)} & \qw \gategroup{1}{2}{4}{2}{.9em}{--} & \targ    & \ctrl{1} & \gate{R_Y(\theta_7)} & \qw \gategroup{1}{3}{4}{6}{.9em}{--} \\
        \lstick{\ket{0}} & \gate{R_Y(\theta_4)} & \qw \gategroup{1}{2}{4}{2}{.9em}{--} & \qw      & \targ    & \gate{R_Y(\theta_8)} & \qw \gategroup{1}{3}{4}{6}{.9em}{--} \\
        }\\
        \text{Layer 1}\hspace{3em} \text{Layer 2}\hspace{5em}
    \end{align*}
    \caption{Illustration of the \texttt{EfficientSU2} ansatz with a linear CNOT entangling layer shown for 4 qubits.}
    \label{fig:ansatz_circuit}
\end{figure}

In order to address the Hamiltonian on IonQ's Aria trapped ion quantum computer, we use VQE~\cite{Peruzzo2014}. Since the solution of the combinatorial optimization problem is given by a computational basis state, we choose an ansatz circuit for VQE that produces only real amplitudes. More specifically, we use a variant of the \texttt{EfficientSU2} ansatz from Qiskit~\cite{Qiskit}, which consists of entangling CNOT gates and parametric $R_Y(\theta) = \exp(-i\theta Y/2)$ rotation gates, where $Y$ is the usual Pauli matrix (see Fig.~\ref{fig:ansatz_circuit} for an illustration).

Moreover, since we are solving a combinatorial optimization problem, we are not directly interested in the properties of the final wave function, and it does not have to be close to the ground state of Eq.~\eqref{eq:Hamiltonian}. We are only concerned with it having a sufficiently large component of a low-energy state of the Hamiltonian, that can be identified with a number of projective measurements that is feasible on the hardware device. By construction, these correspond to good approximations for the solution. Thus, instead of the conventional energy expectation value, we resort to the conditional value at risk (CVaR) as a cost function~\cite{Barkoutsos2020}. Given a random variable $X$ with cumulative density function $F_X$, the CVaR is defined as the conditional expectation over the left $\varepsilon$-tail of the distribution
\begin{equation}
    \text{CVaR}_{\varepsilon}(X) = \mathbb{E}\left[ X | X \leq F^{-1}_X(\varepsilon) \right].
\end{equation}
In the expression above, $\mathbb{E}$ denotes the expected value, $F^{-1}_X$ the inverse of the cumulative density function, and $\varepsilon$ is a parameter in the interval $(0,1]$. In the context of VQE, this corresponds to only considering a fraction $\varepsilon$ of all measurements with the lowest energies for a given set of variationl paramaters $\bm{\theta}$. More specifically, assuming that we perform $K$ measurements resulting in $K$ computational basis states with corresponding energy values $\{E_1, E_2, \cdots ,E_K\}$ sorted in ascending order, the cost function for a given set of parameters is given by
\begin{align}
    C(\bm{\theta}) = \text{CVaR}_{\varepsilon} = \frac{1}{\lceil \varepsilon K \rceil} \sum_{i=1}^{\lceil \varepsilon K \rceil} E_i.
    \label{eq:cvar}
\end{align}
For $\varepsilon=1$ the expression above is nothing but the usual estimate of the energy expectation value with $K$ measurements. The limit $\varepsilon\to 0$ corresponds to just keeping (one of) the measurement(s) with the lowest energy value, which would lead to a cost function that is discontinuous in the variational parameters. Note that the choice of $\varepsilon$ also affects the maximum component of the optimal solution one can expect in the ground state. Since only the fraction  $\varepsilon$ of measurements with the lowest energy contributes to the cost function, there is essentially no reward for increasing the component of the optimal solution in the final state above $\varepsilon$. Thus, in practice one has to choose a reasonable value for $\varepsilon$ which allows for generating a sufficiently large component of the optimal solution, but which is small enough to take advantage of the CVaR.

\section{Results\label{sec:results}}

We examine the performance of the IonQ Aria trapped ion quantum processing unit (QPU) for the FGA problem using several instances of different size. This device uses up to 25 addressable Ytterbium (Yb) ions linearly arranged in an ion trap. Qubit states are implemented by utilizing two states in the ground hyperfine manifold of the Yb ions. Manipulating the qubits in the Aria QPU is done by 355-nm laser pulses, which drive Raman transitions between the qubit states. By configuring these pulses, arbitrary single-qubit gates and M{\o}lmer-S{\o}renson type two-qubit gates~\cite{PhysRevLett.82.1971} can both be realized. As of 2023, the Aria QPU has demonstrated performance at the level of 25 algorithmic qubits~\cite{IonQ-AQ,Lubinski2023}.

In order to mitigate the effect of systematic errors on the Aria QPU, error mitigation via symmetrization is used~\cite{maksymov2023enhancing}. After executing multiple circuit variants with distinct qubit to ion mapping, the measurement statistics is aggregated using component-wise averaging.

For all results presented here, we use VQE with the CVaR as an aggregation function and the ansatz shown in Fig.~\ref{fig:ansatz_circuit}, where we restrict ourselves to two layers of the ansatz circuit which results in $2Q$ variational parameters for $Q$ qubits. To show that current quantum devices are suitable for addressing the FGA problem, we proceed in two steps. First, we demonstrate the full VQE for smaller instances of the problem, where we run the feedback loop between the quantum device and the classical optimizer until convergence. Second, for larger instances we restrict ourselves to performing inference of the result for the optimal parameters on the quantum hardware. This way we are able to address problems on quantum hardware that would require 36 binary decision variables in the QUBO formulation, a problem size for which quadratic assignment problems are in general already hard to solve exactly~\cite{Anstreicher2003,Loiola2007}.

\subsection{VQE with the IonQ Aria trapped-ion quantum computer}
\begin{figure}[htp!]
    \includegraphics[width=1.0\columnwidth]{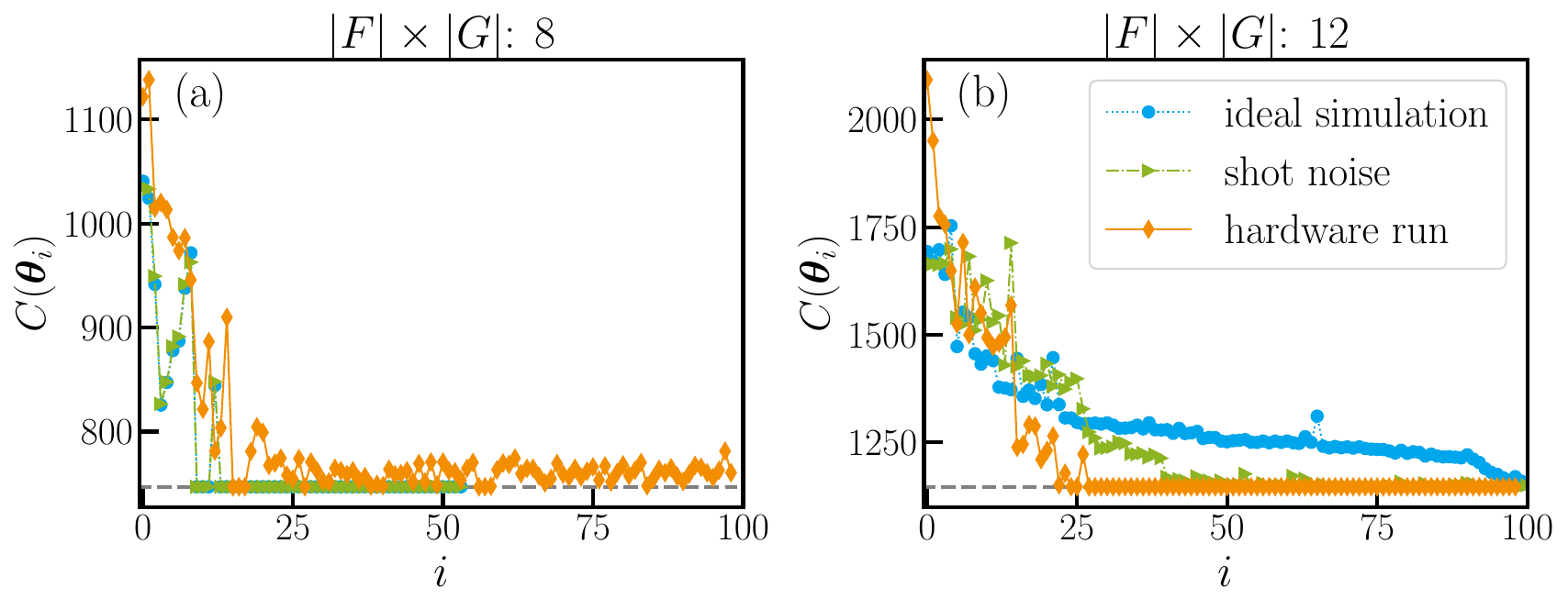}
    \caption{Cost function value versus the iteration for $2$ flights and $4$ gates corresponding to $4$ qubits (a) and for $3$ flights and $4$ gates corresponding to $6$ qubits (b). The blue circles correspond to the results of an ideal simulation with an infinite number of measurements, the green triangles to a simulation with $1000$ measurements, and the orange diamonds to the data obtained on the quantum hardware with 1000 measurements. As a guide for the eye, the data points are connected with lines. The horizontal dashed grey line indicates the minimal value of the cost function.}
    \label{fig:cost_vs_iteration_full_vqe}
\end{figure}

We address two instances of the FGA problem with VQE, one with $2$ flights and $4$ gates which can be encoded into 4 qubits, and one with  $3$ flights and $4$ gates which can be represented with 6 qubits. We use the CVaR as a cost function with a value of $\varepsilon=0.5$ and 1000 measurements for each iteration, which we also refer to as shots. Our results for the cost function during the optimization using the COBYLA optimizer~\cite{Powell1994} are shown in Fig.~\ref{fig:cost_vs_iteration_full_vqe}.

Focusing on the ideal simulation with an infinite number of shots for $Q=4$ qubits in Fig.~\ref{fig:cost_vs_iteration_full_vqe}(a) first, we observe that the cost function does not converge monotonically, as expected for COBLYA. After approximately 10 iterations the minimum of the cost function is found, and continuing the optimization procedure further does not lead to any changes anymore. Thus, we stop the simulation after 51 iterations. Taking into account the shot noise due to finite number of measurements does qualitatively not change the picture, as the green triangles in Fig.~\ref{fig:cost_vs_iteration_full_vqe}(a) reveal. The finite number of shots only leads to slightly different cost function values at the early stage of the optimization, before eventually converging to the minimum. Looking at the data from the quantum device (orange diamonds in Fig.~\ref{fig:cost_vs_iteration_full_vqe}(a)), we see a qualitatively similar behavior than in the classical simulations. The cost function value initially fluctuates before eventually converging close to the global minimum. The values $C({\bm\theta}_i)$ at an early stage differ from the classical simulation, presumably due to the effect of hardware noise that cannot be fully mitigated. Eventually the data also reaches the global minimum after approximately the same number of iterations as the classical simulation. In contrast to the simulated data, the results from the quantum hardware eventually fluctuate around the global minimum of the cost function, most likely due to hardware noise, producing sometimes solutions that are close to the optimal one.

For the larger instance with $Q=6$ qubits Fig.~\ref{fig:cost_vs_iteration_full_vqe}(b), the classically simulated results show a slightly different behavior. The data from the ideal simulation with an infinite number of measurements shows  again some fluctuations at an early stage, and eventually the process converges to the global minimum of the cost function after 100 iterations. Compared to the previous problem with 4 qubits, the convergence in this case is a lot slower. Taking into account the finite number of measurements and the resulting shot noise, the situation improves as the green triangles in  Fig.~\ref{fig:cost_vs_iteration_full_vqe}(b) reveal. While the fluctuations in the cost function seem to be enhanced by the shot noise during early stages of the optimization process, we observe convergence to the global minimum after approximately 40 iterations. The data obtained from the QPU also shows a fast convergence to the global minimum of the cost function. Compared to the previous case with 4 qubits, we do not observe fluctuations of the cost function around the global minimum after 26 iterations and $C({\bm\theta}_i)$ is stable. Interestingly, it seems that the combined hardware and shot noise on the quantum device enhances the convergence, as the data from the QPU reaches the global minimum faster than the classical simulation with shot noise (c.f.\ the green triangles and the orange diamonds in Fig.~\ref{fig:cost_vs_iteration_full_vqe}(b)).

\begin{figure}[htp!]    
    \centering
    \includegraphics[width = 1\columnwidth]{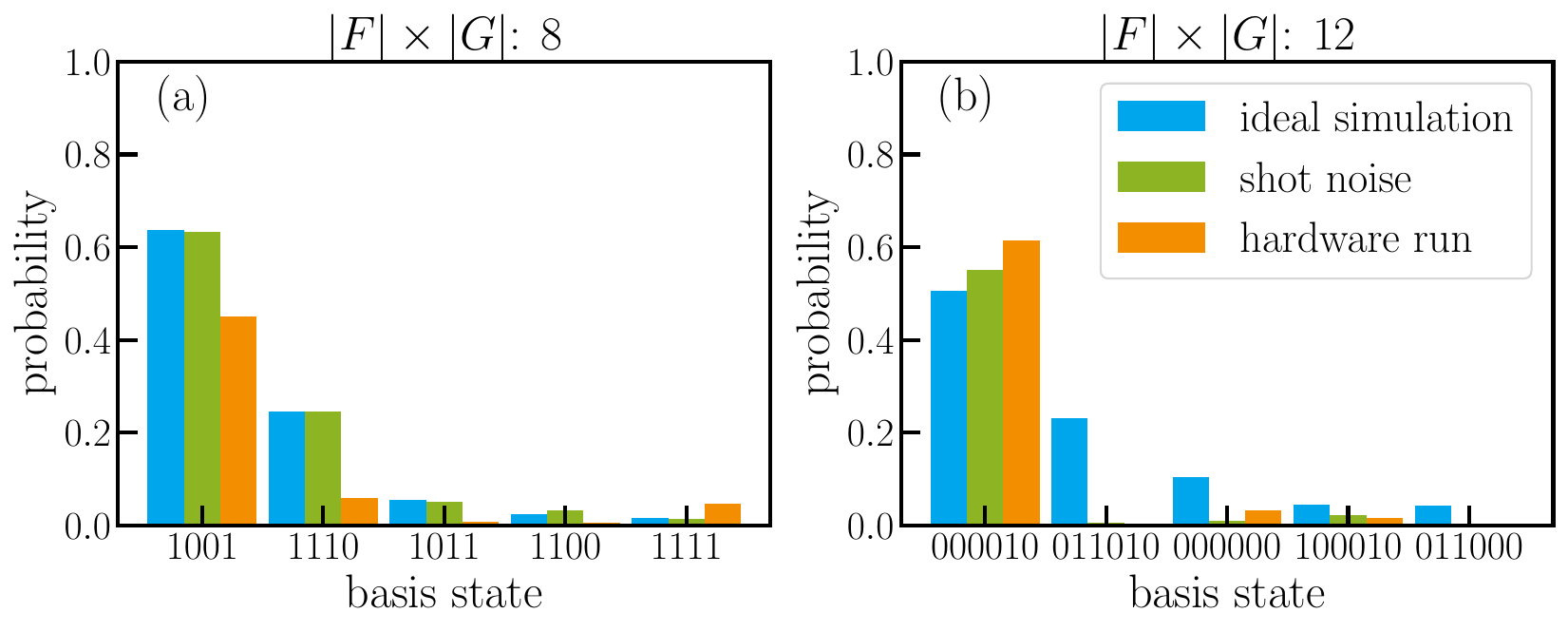}
    \caption{Probability distribution for the 5 most dominant basis states $\ket{z_0\dots z_{Q-1}}$ in the final state obtained from an ideal simulation with infinite shots (blue),  with $1000$ shots (green), and from the measurement on the IonQ Aria trapped ion quantum computer (orange). The different panels correspond to different problem sizes $Q=4$ (a) and $Q=6$ (b).}
    \label{fig:cmp_ideal_hardware_full_vqe}
\end{figure}

To judge the quality of the solution obtained on the quantum device, we investigate the probabilities of measuring the individual basis states and compare those to the ones obtained from the ideal simulation. Again, we consider an ideal simulation with an infinite number of measurements, an ideal simulation with 1000 measurements and the data from the quantum device with 1000 measurements, too. For the latter two cases, we estimate these probabilities by taking the relative frequency of the computational basis states observed. In all cases, we determine these distributions at the end of the final iteration. Figure~\ref{fig:cmp_ideal_hardware_full_vqe} shows the results for the two instances with $4$ and $6$ qubits.
For both cases, we observe that VQE produces a dominant component of the exact solution in the final state. As outlined earlier, the probability of measuring the ground state after running VQE with the CVaR as a cost function should be approximately given by $\varepsilon$ in case the global minimum was reached. For our choice of $\varepsilon=0.5$, we thus expect that the final state produced by VQE has a component ground state of around $50\%$, which is indeed the case, as Fig.~\ref{fig:cmp_ideal_hardware_full_vqe}(a) and Fig.~\ref{fig:cmp_ideal_hardware_full_vqe}(b) show. Focusing on the data for 4 qubits in Fig.~\ref{fig:cmp_ideal_hardware_full_vqe}(a), we observe that the data obtained from the QPU has a slightly smaller ground state component than the final wave function from the classical simulation. This can be explained by the fact that the results on the quantum hardware fluctuate after convergence, and the value of $C({\bm\theta}_i)$ in the final iteration is slightly larger than the global minimum. Nevertheless, the solution produced by VQE on the quantum device has a $40\%$ component of the exact ground state, which is reliably observed with 1000 measurements. 

The histogram for measuring the individual basis states for $Q=6$ in Fig.~\ref{fig:cmp_ideal_hardware_full_vqe}(b) corroborates the picture that in this case noise is beneficial and helps the convergences. While the exact simulation with an infinite number of measurements manages to produce a ground state component of $50\%$, the data obtained from the noisy classical simulation with 1000 shots, which showed faster convergence in the number of iterations, produces an even higher component of the ground state. This effect is even more pronounced for the data from the quantum devices, which shows a probability of measuring the ground state higher than $60\%$, in agreement with the observation that the simulation on the quantum hardware was converging faster.

\subsection{Inference on the IonQ Aria trapped-ion quantum computer}

\begin{figure*}
    \centering
    \includegraphics[width=0.95\textwidth]{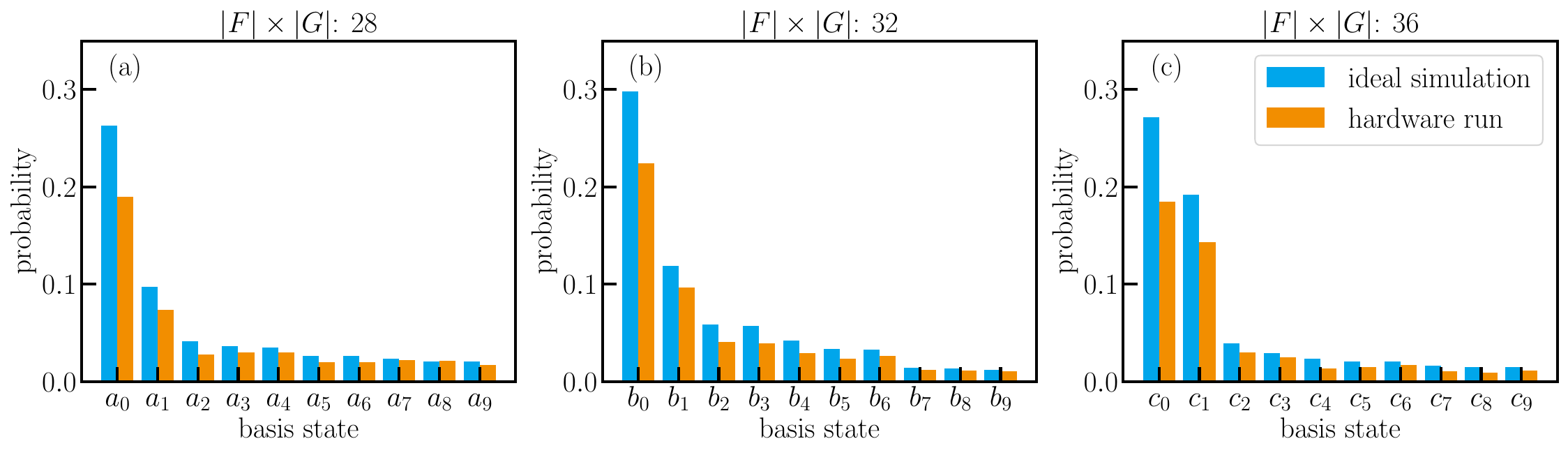}
    \caption{Probability distribution for the 10 most dominant basis states in the final result obtained from an ideal simulation (blue) and from the measurement on the IonQ Aria trapped ion quantum computer (orange). The different panels correspond to different problem sizes $Q=14$ (a), $Q=16$ (b), and $Q=18$ (c).}
    \label{fig:cmp_ideal_hardware_inference}
\end{figure*}

In order to demonstrate the approach can also be scaled up to a larger number of qubits, we perform inference runs for problem instances with a more flights and gates on the quantum hardware. We first carry out a classical simulation of VQE with CVaR as a cost function with $\varepsilon = 0.25$, where we again run a simulation using the state vector, which corresponds to an ideal quantum device without shot noise. The optimization is again done with COBYLA, where we limit the amount of iterations $50\times Q$ to make the data for different problem sizes comparable. Subsequently, we prepare the corresponding wave function on the quantum device and measure the result with 10000 shots to estimate the probabilities of measuring the different basis states. Figure~\ref{fig:cmp_ideal_hardware_inference} shows the experimental results in comparison to the classical simulation for different instances with 14, 16 and 18 qubits.

Comparing the results obtained from an ideal simulation with the ones from the quantum hardware with a finite number of measurements, we see good agreement for the distribution of basis states, indicating that the QPU is able to prepare the state corresponding to the final parameters with good accuracy. Looking at the different panels for different problem sizes, we also observe that increasing the problem size does not seem to lead to a drop in performance, and the results from the hardware agree well with the ideal simulation, even for the largest problem size with $Q=18$ we study. In particular, our data show that the wave function prepared on the quantum device has a dominant component of the ideal solution for all instances, just like the ideal results, which would allows for measuring it with a high probability.

To further characterize results obtained from the IonQ Aria trapped ion processor, we estimate bounds on the fidelity between the hardware results and the ideal solution. To this end let  $\ket{\psi_\text{qc}} = \sum_{i=0}^{2^{Q}-1} a_i \ket{i}$ be the wave function actually prepared on the quantum device and  $\ket{\psi_\text{id}} = \sum_{i=0}^{2^{Q}-1} b_i \ket{i}$ the ideal solution from the classical simulation. Thus, we can obtain an upper bound for the fidelity
\begin{align}
    \begin{aligned}
        F = \left|\ip{\psi_\text{qc}}{\psi_\text{id}}\right|^2  &= \left|\sum_{i=0}^{2^Q-1} a_i^*b_i\right|^2 \leq \left(\sum_{i=0}^{2^Q-1} |a_i| |b_i|\right)^2\\
        &\approx\left(\sum_{i=0}^{2^Q-1} \sqrt{p_i^\text{qc}} \sqrt{p_i^\text{id}}\right)^2
    \end{aligned}
\end{align}
where $p_i^\text{qc}$ and  $p_i^\text{id}$ are the probabilities for measuring the basis state $\ket{i}$ obtained from the quantum device with a finite number of measurements and the ideal simulation respectively. Computing this estimate for the results in Fig.~\ref{fig:cmp_ideal_hardware_inference} we obtain the upper bounds 0.8806, 0.8648, and 0.8448 for the fidelities for 14, 16 and 18 qubits. Note that VQE with the CVaR as a cost function does not require an extremely precise implementation of the quantum state corresponding to the current set of parameters to succeed. As long as the distribution of basis states qualitatively follows the exact one, the method will be able to generate a dominant component of a low-energy state in the final wave function. Thus, quality of the results obtained on quantum device is sufficient to find (a good approximation to) the solution even larger instances of the problem.

\section{Conclusion\label{sec:conclusion}}
In this work we explored the performance of IonQ's Aria trapped ion quantum processor for the FGA problem, a quadratic assignment problem with linear constraints, using VQE with the CVaR as a cost function. Utilizing a qubit efficient encoding for the problem we preformed a full VQE for instances with $4$ and $6$ qubits. In both cases, our data obtained from the QPU exhibit a dominant component of the ground state wave function encoding the optimal solution, and qualitatively in good agreement with the results from classically simulating VQE. Interestingly, for the instance with $6$ qubits we observe faster convergence on the quantum hardware compared to noise-free classical simulation of VQE, indicating that the noise in the quantum device has a positive effect in this case.

In order to demonstrate that the approach can be generalized to larger problem sizes, we performed inference runs for problems with $14$, $16$, and $18$ qubits. Preparing the wave function for the optimal set of parameters obtained from a classical simulation, we showed that the trapped ion quantum processor is able to generate the corresponding wave function with sufficient fidelity to allow for a successful VQE with CVaR as the cost function, and that results from the QPU are in good agreement with the ones from a classical simulation.  Thus, our results provide a promising indication that the FGA problem can be successfully addressed on near-term quantum hardware.

For the future, there are several aspects to explore. On the one hand, while it was shown that noise can be detrimental in variational quantum algorithms~\cite{Wang2021}, there has been some theoretical work suggesting that under certain circumstances there can be a positive effect~\cite{Liu2022Noise}. Our results for small instances indicate that this might also be the case for the FGA problem and we plan to study the effect of noise on the problem in more detail. On the other hand, it will be interesting to study the scaling of resources and the performance for larger problem sizes. Our results indicate that there is no strong dependence on the number of qubits, which we plan to explore in future simulations on quantum hardware.

\begin{acknowledgements}
This work is funded by the European Union’s Horizon Europe Framework Programme (HORIZON) under the ERA Chair scheme with grant agreement no.\ 101087126.
This work is supported with funds from the Ministry of Science, Research and Culture of the State of Brandenburg within the Centre for Quantum Technologies and Applications (CQTA). 
\begin{center}
    \includegraphics[width = 0.08\textwidth]{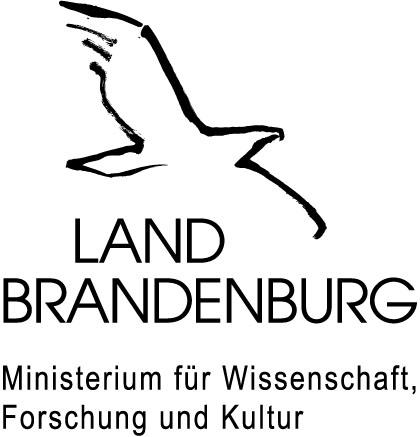}
\end{center}

\end{acknowledgements}

\bibliography{bibliography}
\end{document}